\pdfoutput=1
\documentclass[aip,apl,amsmath,amssymb,
reprint,
]{revtex4-1}

\usepackage{graphicx}
\usepackage{dcolumn}
\usepackage{bm}

\begin{document}

\title{Tunneling spectroscopy of superconducting MoN and NbTiN grown by atomic layer deposition}

\newcommand{\argonne}{\affiliation{Materials Science Division, Argonne National Laboratory, Argonne, Illinois 60439, USA}}
\newcommand{\iit}{\affiliation{Department of Physics, Illinois Institute of Technology, Chicago, Illinois 60616,USA}}
\newcommand{\insitu}{{\itshape in situ }}
\newcommand{\nbtin}{Nb$_{0.8}$Ti$_{0.2}$N }
\newcommand{\alo}{Al$_{2}$O$_{3}$ }

\author{Nickolas R. Groll}
\email{ngroll@anl.gov}
\argonne
\author{Jeffrey A. Klug}
\argonne
\author{Chaoyue Cao}
\argonne
\iit
\author{Serdar Altin}
\affiliation{Fen Edebiyat Fakultesi, Fizik Bolumu, Inonu Universitesi, 44280 Malatya, Turkey}
\author{Helmut Claus}
\argonne
\author{Nicholas G. Becker}
\argonne
\iit
\author{John F. Zasadzinski}
\argonne
\iit
\author{Michael J. Pellin}
\argonne
\author{Thomas Proslier}
\email{proslier@anl.gov}
\argonne

\date{\today}

\begin{abstract}
A tunneling spectroscopy study is presented of superconducting MoN and \nbtin thin films grown by atomic layer deposition (ALD). The films exhibited a superconducting gap of 2~meV and 2.4~meV respectively with a corresponding critical temperature of 11.5K and 13.4K, among the highest reported $T_c$ values achieved by the ALD technique.  Tunnel junctions were obtained using a mechanical contact method with a Au tip.  While the native oxides of these films provided poor tunnel barriers, high quality tunnel junctions with low zero bias conductance (below $\sim$10\%) were obtained using an artificial tunnel barrier of \alo on the film's surface grown $ex$ $situ$ by ALD.  We find a large critical current density on the order of $4\times 10^6$A/cm$^2$ at $T=0.8T_c$ for a 60nm MoN film and demonstrate conformal coating capabilities of ALD onto high aspect ratio geometries.  These results suggest the ALD technique offers significant promise for thin film superconducting device applications.
\end{abstract}

\maketitle

Superconducting nitride alloys make excellent candidates for device applications due to their relatively high critical temperature and stability at ambient atmosphere. These characteristics suggest the opportunity for increased performance and efficiency over commonly used niobium. However, challenges and limitations in common fabrication techniques can reduce the application possibilities. Reactive magnetron sputtering is the most commonly used deposition technique that has produced high quality superconducting thin films of niobium based nitrides.\cite{vanDoverAPL1982,BaconJAP1983,ThakoorJAP1985} While molybdenum based superconducting nitrides have also been achieved through sputtering\cite{IharaPRB1985} or chemical solution\cite{ZhangJACS2011} techniques, they require a high temperature ($\geq$700$^{\circ}$C) and/or high pressure annealing process to achieve a critical temperature greater than that of niobium. Various other techniques such as chemical vapor deposition,\cite{ShiPRB1988,*GaninJSSC2006} pulsed laser deposition\cite{InumaruCM2005,*InumaruPRB2006} and molecular beam epitaxy\cite{InumaruASS2006} have produced superconducting molybdenum nitrides, but have failed to even reach 9K. Sputtering methods are generally limited to planar geometries making it very difficult for complex geometry applications such as superconducting radio-frequency (SRF) cavities or superconducting magnets. Additionally, for applications requiring thin pin-hole free insulating barriers such as superconductor-insulator-superconductor (SIS) tunnel junction mixers for terahertz frequencies,\cite{UzawaAPL1998,*GundlachSST2000,*IosadASIT2001} sputtering processes can be quite challenging.

Atomic layer deposition (ALD) is a highly scalable technique that utilizes sequential self-limiting surface chemical reactions to deposit material in a layer-by-layer fashion.\cite{Ritalam.2001,puurunenjap2005} ALD provides several advantages over traditional growth methods, including atomic-scale uniformity over large areas, unmatched conformity over complex-shaped substrates, and deposition temperatures well below those often required by other techniques. ALD has previously demonstrated the ability to grow superconducting films\cite{KlugJPCC2011,*ProslierJPCC2011} and nitride based thin films in particular.\cite{hiltunenTSF1988} Currently, reports on the superconducting properties of ALD films with Tc's above 9K have been limited to standard resistance and susceptibility measurements. The bulk of the letter will focus on the characterization of the superconducting properties of \nbtin and MoN thin films grown by ALD through the use of electron tunneling spectroscopy. We demonstrate the ability to create artificial tunnel barriers with ALD, with potential applications to devices based on SIS tunnel junctions. Finally, critical current density $J_c$ measurements are performed for MoN.  The high $J_c$ value along with the conformal deposition capabilities of ALD thin films are relevant to complex geometry device applications.

\begin{figure*}[ht]
\centering
\includegraphics[width=\textwidth]{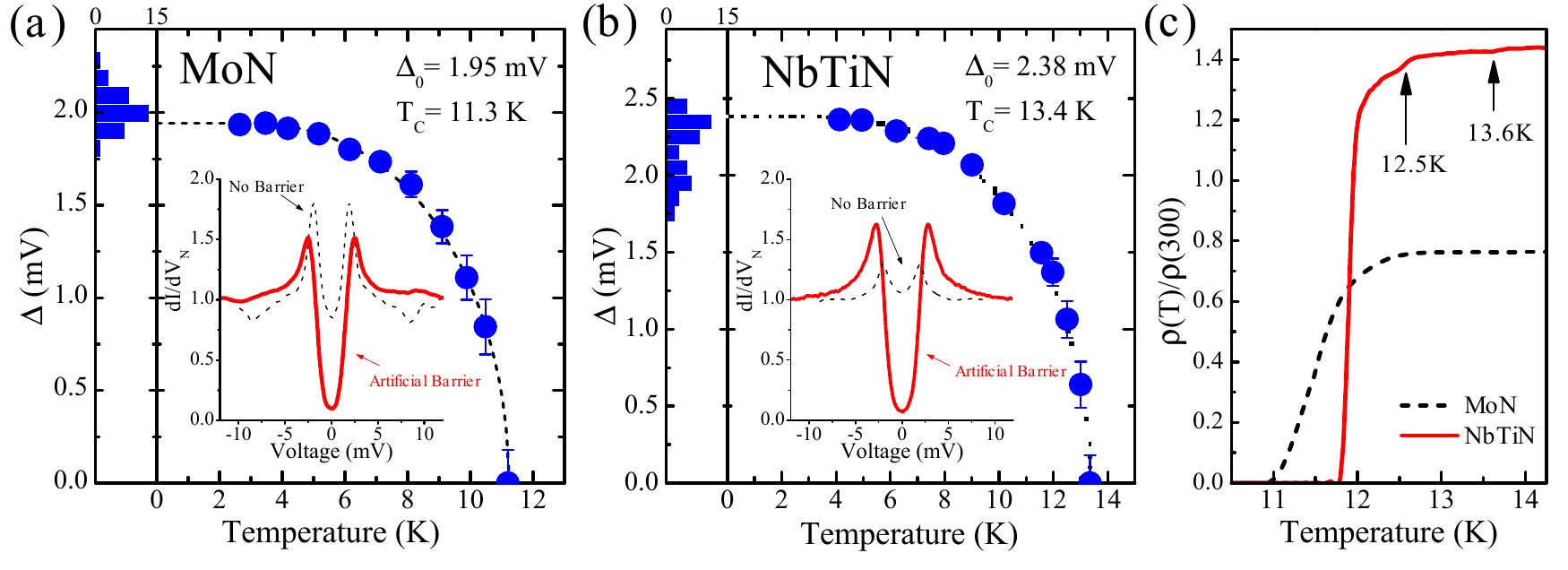}
\caption{(Color online) Summary of the PCT data obtained for the MoN film (a) and NbTiN film (b). The insets of (a) and (b) show the normalized $dI/dV$ curves for each film before (dashed curves) and after (solid curves) the deposition of an artificial tunnel barrier. The histogram of gap values obtained for each film is shown on the left of each panel for approximately 35-40 junctions measured. The four wire transport measurements are shown in (c) where we find a single transition for the MoN film (dashed) and multiple for the NbTiN film (solid red).}
\label{fig:PCT}
\end{figure*}

Point contact tunneling (PCT) spectroscopy\cite{OzyuzerC1998} is a powerful tool for studying superconductors as the measure of the current $I_{NS}(V)$ flowing between the normal metal $N$ (tip) and the superconductor $S$ (sample) under a difference of potential $V$ allows for a direct measurement of the surface superconducting properties. Using the modified Blonder-Tinkham-Klapwijk (BTK) theory\cite{BlonderPRB1982, *DynesPRL1978} we are able to extract the superconducting gap $\Delta$, barrier strength $Z$ and phenomenological quasi-particle lifetime broadening parameter $\Gamma$ from fits of the differential conductance $dI(V)/dV$. In the tunnel regime ($Z\gg1$), the differential conductance is directly proportional to the superconducting density of states, $N_S (E)$, through the relation
\begin{equation}
\frac{dI_{NS}}{dV}\propto \int^{\infty}_{\infty}N_{S}(E)\left[ -\frac{\partial f(E+eV)}{\partial (eV)} \right]dE,
\end{equation}
where $f(E)$ is the Fermi function. In our setup, the point contact junction is made with the sample by approaching the surface with a sharpened gold tip to create a SIN (superconductor-insulator-normal) junction where an oxide layer serves as the insulator. The tip-sample separation is controlled through the use of a differential screw while mechanical hysteresis of the tip motion allows for probing slightly different regions of the surface and obtaining statistics of the sample's superconducting properties. A homemade analog sweep generator was combined with a lock-in amplifier to simultaneously measure the $IV$ and $dI/dV$ spectra of the junction. 

In general, we use the formation of a native oxide barrier at the surface of the sample as a natural barrier to achieve a SIN tunnel junction with the point contact technique. However, the nitride superconductors measured in this work do not readily form a sufficient insulating oxide layer, requiring the introduction of an artificial barrier in order to reach the tunneling regime. Aluminum oxide was selected as the tunneling barrier\cite{BelkinAPL2010,*ProslierAPL2008} for several reasons. From a point contact standpoint, Al$_2$O$_3$ is a well studied and commonly used oxide barrier that has produced high quality planar tunnel junctions.\cite{ZasadzinskiPRB1982} Additionally, amorphous Al$_2$O$_3$ can be grown at temperatures as low as 33$^{\circ}$C using ALD,\cite{GronerCM2004} making it a natural choice to achieve a pin-hole free insulating layer with minimal possibility of damaging or diffusing into delicate substrates.  For this study, the artificial tunnel barrier was not deposited \insitu in order to establish the efficacy of the artificial barrier. After establishing a baseline for PCT measurements of the bare films, a thin aluminum oxide barrier (2-3~nm) was deposited at 85$^{\circ}$C using Al(CH$_3$)$_3$ and H$_2$O precursors in a commercial Savannah S100 ALD system.

The nitride films were deposited on various substrates at 450$^{\circ}$C in a custom-built hot-walled viscous flow reactor with details of the deposition described elsewhere.\cite{KlugAPL2013,*ElamRSI2002} A 60nm thick MoN film was deposited using a binary process using MoCl$_5$ and NH$_3$ precursors and a 45nm thick film of \nbtin was deposited using a NbN:TiN ALD cycle ratio of 1:2. In both cases, growth of the superconducting layer was preceded \insitu by deposition of a 10-20 nm thick AlN buffer layer in order to improve the film nucleation and crystallinity. AlN was grown by ALD using AlCl$_3$ and NH$_3$ precursors at 450$^{\circ}$C.

Approximately 35-40 junctions were measured for each sample after the deposition of aluminum oxide to obtain reasonable statistics of the superconducting parameters. For both films, we obtain remarkable improvement in the zero bias conductance (ZBC) after the deposition of the artificial barrier and are able to measure high quality tunnel junctions. Prior to the aluminum oxide deposition, we did not reach the tunnel regime for either film as seen by the dotted curves in the insets of Fig.~\ref{fig:PCT}(a-b). Here the ZBC is nearly equal to that at high bias indicating that we are somewhere between tunneling and Andreev reflection where $Z<1$ for both films. Looking now at the solid curves shown in the insets of Fig.~\ref{fig:PCT}(a-b), one sees the artificial barrier greatly decreases the ZBC to approximately (or below) 10\% of the high bias value allowing us to reach the tunneling regime where $Z\rightarrow\infty$. This trend has been measured reproducibly for all the junctions and is a clear indication of the the effectiveness and homogeneity of an ALD deposited artificial tunnel barrier. 

\begin{figure}[ht]
\centering
\includegraphics[width=0.5\textwidth]{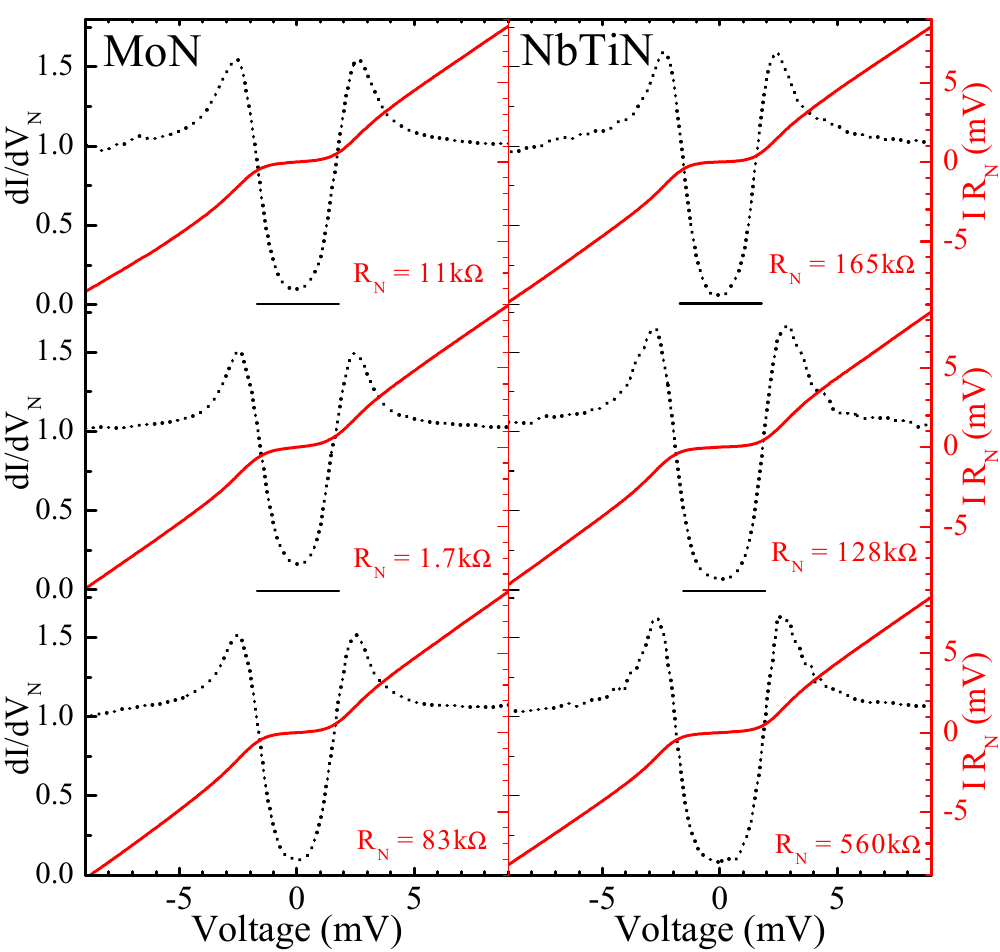}
\caption{(Color online) Typical $IV$ and normalized $dI/dV$ spectra for MoN (left) and NbTiN (right) obtained from point contact tunnel junctions. The curves are shifted vertically for clarity and the $IV$ curves are scaled according to their high bias resistance. Both films exhibit a low ZBC as well as fairly sharp peak features.}
\label{fig:IV}
\end{figure}

A summary of the PCT measurements for the MoN film are shown in Fig.~\ref{fig:PCT}(a). On the left, a histogram of the gap statistics show a single gap value of 2.02$\pm$0.25meV and indicates that we have homogeneous surface superconducting properties. In the main plot of of Fig.~\ref{fig:PCT}(a) is shown the measured local gap temperature dependence of a tunnel junction and the fit results according to the BCS temperature dependence (dashed curve) for an s-wave superconductor. Here is found a local gap value of $1.95\pm0.03$meV and a junction $T_c$ of 11.3$\pm 0.05$K resulting in the ratio $2\Delta/k_B T_c$=4. The local $T_c$ is consistent with the bulk $T_c$ of 11.5$\pm$0.7K measured by standard four wire transport (dashed curve in Fig.~\ref{fig:PCT}(c)) and using the bulk values, the ratio $2\Delta/k_B T_c$=4.1. In the left panel of Fig.~\ref{fig:IV} are shown $IV$ (solid red) and normalized $dI/dV$ (dashed) spectra obtained for typical junctions of the MoN film. We consistently find tunnel junctions having a ZBC in the range of 10-20\% with reasonably sharp energy gap features. This demonstrates not only the feasibility of the artificial barrier, but also that bulk superconductivity appears at the film surface, without any evidence of proximity effects or other surface degradation. 

The same series of measurements were conducted on the NbTiN sample for which a summary of the results is displayed in Fig.~\ref{fig:PCT}(b). In the left plot, the histogram of the measured gap values show that there appear to be two peaks in the distribution, one centered at 2.32$\pm 0.23$meV and the other at 1.99$\pm 0.33$meV. We conducted a local measure of $T_c$ (main plot) by PCT near the maximum in the gap distribution and find a $T_c$ of 13.4$\pm$0.05K and a gap of 2.38$\pm$0.03meV resulting in a $2\Delta/k_B T_c$ ratio of 4.1. This result is consistent with the transport measurement where we also have multiple transitions (solid red curve in Fig.~\ref{fig:PCT}(c)), the highest at 13.6$\pm$0.15K and the lowest at 11.9$\pm$0.13K. The gap statistics and the transport measurements indicate the presence of multiple phases in the Nb$_x$Ti$_{1-x}$N film. The large gap (2.4 meV) with a $T_c$ of 13.4K is measured with a higher probability by PCT and yet appear only weakly in the transport measurement. This observation suggests that the corresponding phase is localized mostly at the surface of the film as disconnected islands, whereas the smaller gap (2 meV) would correspond to the bulk of the film with a $T_c$ of 11.9K resulting in a ratio $2\Delta/k_B T_c$ = 3.9. These inhomogeneities could be due to the more complicated chemistry involved in the deposition leading to a formation of islands with slightly different compositions of the Nb$_x$Ti$_{1-x}$N film. Like the MoN film, the NbTiN film also exhibited consistently high quality tunnel junctions with single gap features. In the right panel of Fig.~\ref{fig:IV}, typical $IV$ and normalized $dI/dV$ curves are shown having low ZBC ($\sim$10\%) and sharp features that could not be obtained relying on a native oxide formation.

For both films studied in this work, we are able to achieve high quality tunnel junctions on surfaces that do not readily form a strong oxide using an ALD deposited insulator. These results provide strong evidence for the feasibility of using ALD to both grow high quality superconducting films as well as create artificial barriers for improved tunneling measurements. ALD is a strong candidate that could be used instead of or in synergy with conventional deposition techniques for superconducting device applications.

\begin{figure}
\centering
\includegraphics[width=0.5\textwidth]{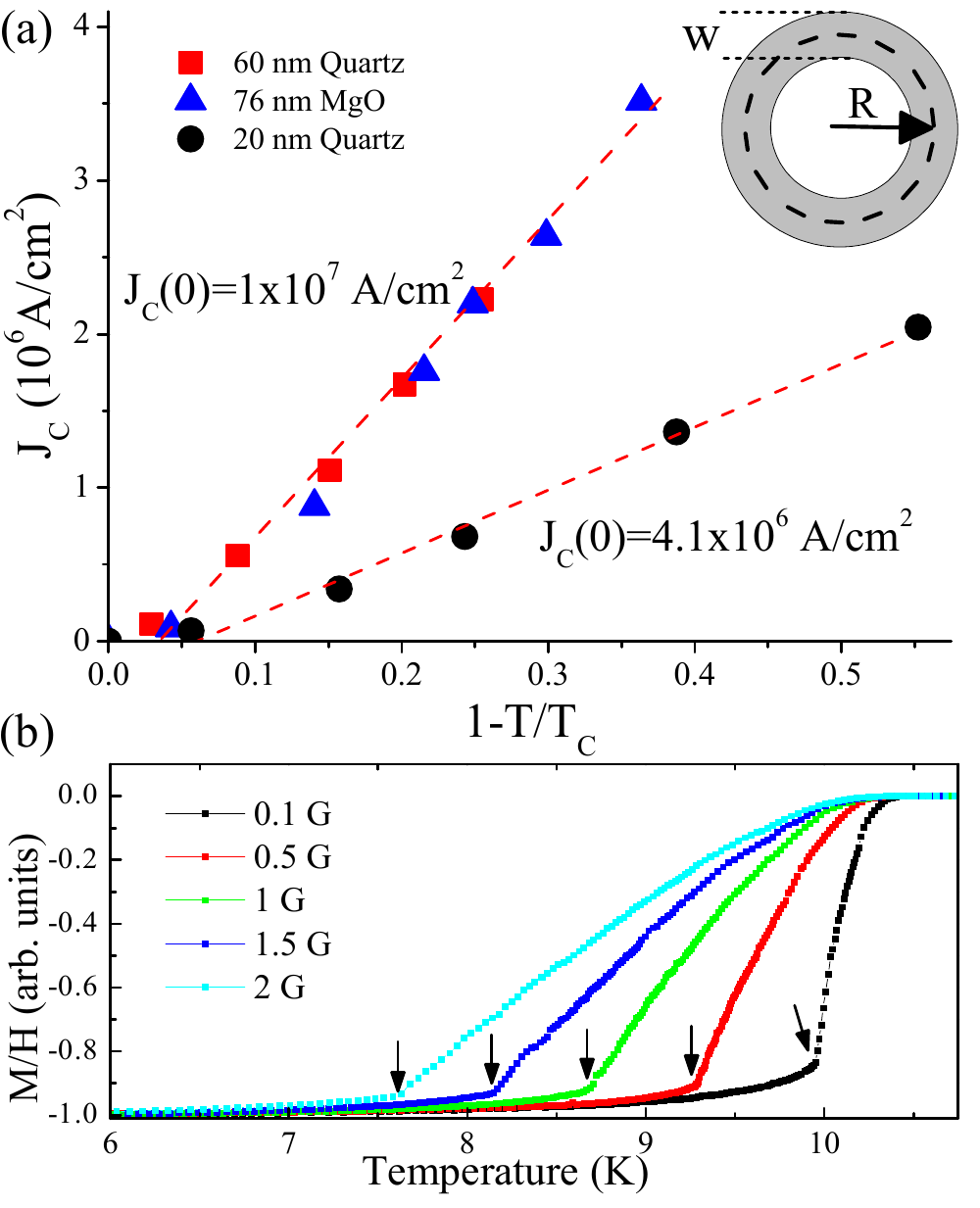}
\caption{(Color online) Critical current density measurements for MoN rings plotted as a function of $T/T_c$ for several thicknesses and substrates (a). The 60nm and 70nm films show remarkably high critical current densities with the 20nm film only slightly lower. (b) Several magnetization curves for the 60nm MoN film where arrows (kink in curve) indicate the temperature at which the field has fully penetrated the ring. }
\label{fig:JC}
\end{figure}

In devices with complex geometries such as SRF cavities and superconducting magnets, the single largest hurdle is the cost incurred using bulk superconducting materials. One straightforward way to drastically reduce the cost of such devices would be to coat inexpensive non-superconducting materials with superconducting films to achieve the same or even improved performance. To investigate the feasibility of ALD deposited superconducting films in these applications, we characterized the critical current density, $J_c(T)\propto1-T/T_c$, and growth characteristics on a high aspect ratio geometry ($\sim$100:1). We limit this proof-of-principle study only to the MoN thin films due to the simpler nature of the MoN deposition chemistry. Following the inductive technique described in reference~\citenum{HerzogPRB1997,*ClausPRB2001}, the $J_c(T)$ was extracted from the magnetic response of thin film rings of MoN. Films of different thicknesses, $t$, were deposited on MgO and fused quartz substrates that were then ion-milled into a ring geometry (see Fig.~\ref{fig:JC}(a)) having a central radius, $R$=1.5mm and width, $w$=150$\mu$m. The induced current $I_{ind}$ in the ring is given by:
\begin{equation}
I_{ind} =  \frac{-\pi R}{\ln{(8R/w)}-1/2}\times H_a
\end{equation}
where $H_a$ is the applied field. In our geometry we have the simple relation $I_{ind}=1.05\times H_a$, where $H_a$ is in mT. Using a SQUID (Superconducting QUantum Interference Device) magnetometer, we measured the temperature dependence of the magnetization curves for three different film thicknesses under several applied magnetic fields. At a given temperature, when the critical-current $I_c(T)=J_c(T) w t$ is reached, the applied field fully penetrates the ring resulting in a kink in the magnetization curves as indicated by the arrows in Fig.~\ref{fig:JC}(b). From this, we extracted the critical current density as a function of temperature shown in Fig.~\ref{fig:JC}(a). The results show a remarkable $J_c(0)$=$1\times 10^7$A/cm$^2$ for a 60nm thin film and only slightly smaller at $4.1\times 10^6$A/cm$^2$ for a 20nm film. If we consider a copper wire with a diameter of 50$\mu$m coated with 60nm of MoN, the critical current would be nearly 1 amp at 2K. Scaling up the number of copper wires could then offer the possibility of tremendous current carrying capabilities for superconducting magnet applications.

\begin{figure}
\centering
\includegraphics[width=0.5\textwidth]{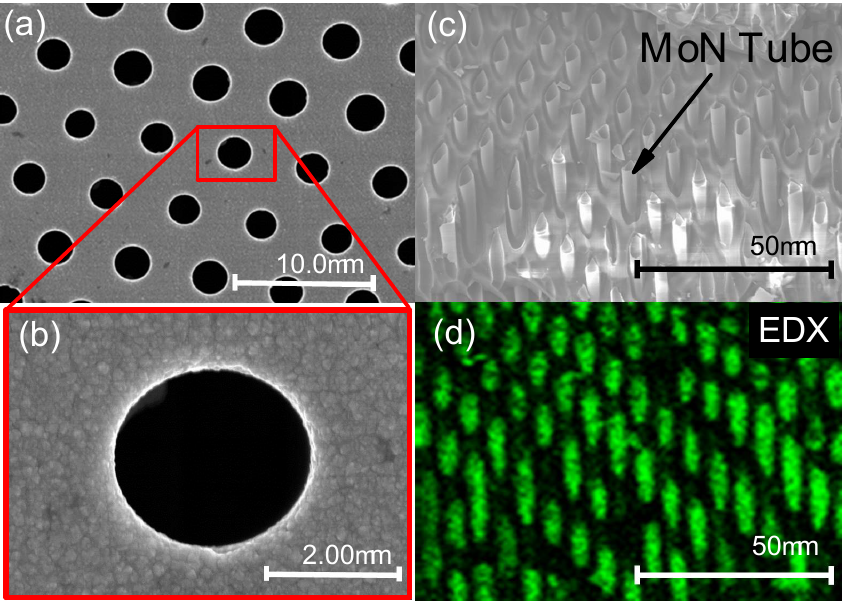}
\caption{(Color online) SEM image of MoN film deposited on a borosilicate substrate (a) containing a few micron diameter capillary tubes with a zoomed view shown in (b). A cross-sectional view approximately 150$\mu$m below the substrate surface is shown in (c) where the tubes sticking out of the capillaries are the deposited MoN. An elemental analysis using EDX (Mo L$\alpha$ peak represented by green in (d)) shows a N/Mo ratio of 1.03 for these tubes, confirmed the composition to be stoichiometric MoN. }
\label{fig:SEM}
\end{figure}

In order to illustrate the conformal deposition capabilities of ALD for potential applications, we deposited 76nm of MoN onto a borosilicate substrate containing a hexagonal pattern of capillary tubes a few microns in diameter and analyzed the deposition using a scanning electron microscope (SEM) equipped with energy dispersive X-ray (EDX) spectroscopy. The measurements reveal a uniform film coating of the surface with no evidence of plugged capillaries (Fig.\ref{fig:SEM}(a-b)), indicative of a self-limiting layer-by-layer deposition. A cross section of the sample was obtained by breaking the membrane in half and observing the penetration of the deposition into the capillary tubes. In Fig.~\ref{fig:SEM}(c), MoN tubes can be seen sticking out of the broken capillaries approximately 150$\mu$m below the substrate surface (1mm total substrate thickness). We performed an elemental analysis using EDX where we found these tubes to be composed of molybdenum and nitrogen with a N/Mo ratio of 1.03, confirming a stoichiometric deposition of MoN. 

In summary, we have measured high quality tunnel junctions on MoN and NbTiN thin films grown by atomic layer deposition with the application of an artificial tunnel barrier. These films exhibit larger $\Delta$ and $T_c$ than that of conventionally used niobium, offering the possibility of improved performance over current niobium-based devices. We have shown that ALD has the capability of coating complex geometries that can have tremendous advantages over conventional deposition techniques. ALD deposited MoN films exhibit a very large critical current density that can be exploited for superconducting magnet architectures. Currently, we are working to further optimize the deposition parameters to increase $T_c$ and incorporate these films into device applications. We are investigating the use of ALD to create \insitu SIS tunnel junctions as well as coating normal metals for complex geometry devices.

The authors would like to thank Incom for providing us with the borosilicate substrates. This work was funded by American Recovery and Reinvestment Act (ARRA) thru the US Department of Energy, Office of High Energy Physics Department of Science to Argonne National Laboratory. The author Serdar Altin would like to thank TU\.{B}ITAK-B\.{I}DEP for financial support during this study. 
Use of the Center for Nanoscale Materials and the Electron Microscopy Center at Argonne National Laboratory were supported by the U.S. Department of Energy, Office of Science under Contract No. DE-AC02-06CH11357.

\end{document}